\documentclass[aps,pre,twocolumn,footinbib,noeprint,longbibliography]{revtex4-1}

\usepackage[utf8]{inputenc}
\usepackage[T1]{fontenc}
\usepackage{amsmath}\usepackage{amssymb}
\usepackage{siunitx}
\usepackage{graphicx}
\usepackage{booktabs}
\usepackage{hyperref}
\usepackage{xcolor}

\hypersetup{%
  colorlinks=true,
  linkcolor=blue,
  citecolor=blue,
  urlcolor=blue,
  linkbordercolor=red,
  pdfborderstyle={/S/U/W 1}
}

\usepackage{tikz}
\newcommand{\revise}[1]{{#1}}

\begin{document}

\title{Dynamics of condensate formation in stochastic transport with\\
pair-factorized steady states: Nucleation and coarsening time scales}

\author{Hannes Nagel}
\email{Hannes.Nagel@itp.uni-leipzig.de}
\affiliation{Institut für Theoretische Physik, Universität Leipzig,
  Postfach 100\,920, 04009 Leipzig, Germany}

\author{Wolfhard Janke}
\email{Wolfhard.Janke@itp.uni-leipzig.de}
\affiliation{Institut für Theoretische Physik, Universität Leipzig,
  Postfach 100\,920, 04009 Leipzig, Germany}

\begin{abstract} Driven diffusive systems such as the zero-range
  process (ZRP) and the pair-factorized steady states (PFSS)
  stochastic transport process are versatile tools that lend
  themselves to the study of transport phenomena on a generic
  level. While their mathematical structure is simple enough to allow
  significant analytical treatment, they offer a variety of
  interesting phenomena. With appropriate dynamics, the ZRP and PFSS
  models feature a condensation transition where for a supercritical
  density the translational symmetry breaks spontaneously and excess
  particles form a single-site or spatially extended condensate,
  respectively. In this paper we numerically study the typical time
  scales of the two stages of this condensation process: Nucleation
  and coarsening. Nucleation is the first stage of condensation where
  the bulk system relaxes to its stationary distribution and droplet
  nuclei form in the system. These droplets then gradually grow or
  evaporate in the coarsening regime to finally coalesce in a single
  condensate when the system finally relaxes to the stationary state.

  We use the ZRP condensation model to discuss the choice of the
  estimation method for the nucleation time scale and present scaling
  exponents for the ZRP and PFSS condensation models with respect to
  the choice of the typical droplet \revise{nuclei} mass. We then
  proceed to present scaling exponents in the coarsening regime of the
  ZRP for partially asymmetric dynamics and the PFSS model for
  symmetric and asymmetric dynamics.
\end{abstract}

\maketitle

\section{Introduction}
\label{sec:introduction}

Stochastic transport processes have been studied for a long time to
understand fundamental principles in various physical and non-physical
systems on a wide range of length-scales. Originally introduced to
study properties of interacting Markov processes by
Spitzer~\cite{Spitzer1970}, they were found to be versatile enough to
provide mappings or models of more realistic systems.  For instance,
extended versions of the asymmetric exclusion process and related
models have been used to model intracellular locomotion of molecular
motors~\cite{Garai2009} as well as vehicular and other
traffic~\cite{Chowdhury2000Review, Chowdhury2005,
  Schadschneider2011}. To us, another interesting field of application
is the modeling and understanding of general condensation phenomena
that are observed in a broad range of physical systems, such as
colloidal and granular systems~\cite{Shim2004}, the formation of
breath figures~\cite{Beysens1986}, \revise{the condensation on
  inhomogeneous networks~\cite{Bogacz2007, Waclaw2007, Waclaw2008b},}
the aggregation of links in networks~\cite{Kwon2011} as well as in a
variety of other contexts~\cite{Evans2005, Chowdhury2000Review,
  Schadschneider2011}. Some stochastic transport processes, in general
driven far from equilibrium, lend themselves to this domain due to
their relative simplicity, the occurrence of phase separation and the
resulting rich phase structures being present already in one
dimension~\cite{Evans2000}.  For example, the asymmetric simple
exclusion process (ASEP) features a phase diagram with a maximum
current, as well as a low-density and a high-density phase induced by
externally driven particle exchange at the system
boundaries~\cite{Derrida1992, Derrida1993, Kolomeisky1998,
  Derrida1998,Blythe2004}. Another well-known example for the
occurrence of phase separation in a one-dimensional stochastic
transport model is the zero-range process
(ZRP)~\revise{\cite{Spitzer1970, OLoan1998, Evans2000, Evans2015}},
where particles on the same site interact non-exclusively: With
appropriate dynamics, this model features a condensation transition
with a coexisting localized particle condensate and a ``fluid''
background or bulk phase. For sufficiently high particle density
$\rho$, the translational symmetry breaks down spontaneously and a
condensate containing a finite fraction of particles emerges at a
single site in the system while the remaining particles form the bulk
of the system.  This condensation model within the ZRP has been
extensively discussed in the literature, so that a wide host of
properties is well known and many related models were proposed and
studied.  For example, a model with nearest-neighbour interactions
proposed by Evans et al.~\cite{Evans2006} features formation of
spatially extended condensates, a property the ZRP lacks due to its
interaction range being zero.  In this model, the stationary state
factorizes over lattice bonds instead of sites as for the ZRP, which
is why it is often referred to as the pair-factorized steady states
(PFSS) model, mainly to distinguish from the fully factorized steady
state of the ZRP.  Many stationary properties of the condensed state
in this family of models such as the condensate's shape and length
scale~\cite{Waclaw2009c, Waclaw2009b, Ehrenpreis2014},
mobility~\cite{Godreche2005, Hirschberg2012}, metastability of the
condensate~\cite{Chleboun2010} or boundary induced phase
separation~\cite{Levine2005,Nagel2015a,Nagel2016a} were studied.

Besides these interesting stationary properties of the condensed state
of these models, the dynamics that lead to it are of interest to the
study of general condensation phenomena.
In this paper we will discuss the dynamics of the condensation process
for the PFSS model as well as the ZRP. Specifically we will determine
and discuss the time scales of the two main steps of the relaxation
process from a random initial state to the final stationary
distribution where a single condensate contains the entire excess
particles of the system. These general steps are the \emph{nucleation}
and \emph{coarsening} processes \revise{as in the well-known picture
  of Ostwald ripening~\cite{Ostwald1897}. Classical theories of
  precipitation in this context are the Szilard--Farkas and
  Becker--D{\"o}ring models~\cite{Farkas1927,Becker1935} for
  nucleation and the Lifshitz--Slyozov--Wagner
  model~\cite{Lifshitz1961} in the coarsening regime.}
Nucleation is the initial diffusive accumulation of several particles
to small droplets that are larger than the typical fluctuations in the
critical background of the relaxed system. Coarsening refers to the
growth of some droplets and the evaporation of smaller ones until only
a single large droplet persists and the steady state is reached. While
the steady state might be in non-equilibrium due to an external drive
and particle transport, this process irreversibly relaxes the system
from a non-equilibrium state towards the stationary distribution. The
time scale of these processes refers to the dependence of the
nucleation and coarsening times to the size of the system.

For the ZRP, the nucleation and coarsening steps and the corresponding
time scales have been treated analytically for symmetric as well as
totally asymmetric dynamics by Grosskinsky
et~al.~\cite{Grosskinsky2003}. In an earlier study the coarsening time
scale of a finite ZRP has been determined by interpreting the size of
the largest condensate as a bounded and biased random walk and
calculating the first-passage time when the largest condensate
absorbed any particles of a smaller
condensate~\cite{Godreche2003,Godreche2005}.

In this article we will extend these considerations for the ZRP to the
generalized PFSS condensation model proposed by Evans
et~al.~\cite{Evans2006} featuring short-range interactions. While this
allows us to create more realistic models of transport phenomena, its
mathematical structure remains simple and similar to the ZRP. Most
importantly here is, that the introduction of nearest-neighbour
interactions allows the emergence of spatially extended
condensates. In fact, variations of the hopping dynamics have been
proposed that lead to interesting properties of the condensate, such
as a tunable envelope shape and scaling of the condensate
width~\cite{Waclaw2009c,Waclaw2009b,Ehrenpreis2014}.  In a recent
contribution we discussed another variant as a toy model in the
context of thin film growth~\cite{Ochab2015}, were the system starts
empty and condensation gradually takes place as particles are
deposited from the outside. We are therefore interested in comparing
the nucleation and coarsening processes of extended droplets in this
model to those observed in the ZRP and specifically determine the
respective time scales.

We will continue as follows.  We first introduce the ZRP and PFSS
models and discuss the expected time scales in Sec.~\ref{sec:models}.
In the following Sec.~\ref{sec:methods}, we discuss the numerical
methods employed to simulate the dynamics and track the condensation
process.  We will further describe our considered methods to compute
the time scale exponents from the trajectories.  In
Secs.~\ref{sec:nucleation} and \ref{sec:coarsening}, we compare our
results for the condensation time scales of the ZRP and PFSS processes
respectively.  Finally, the last Sec.~\ref{sec:summary} contains the
conclusions of our work.

\section{Models}
\label{sec:models}

We consider condensation phenomena in two related models of stochastic
transport processes: The zero-range process (ZRP) and the
pair-factorized steady states (PFSS) model. The ZRP is a minimal model
of a particle-hopping stochastic transport process with local
interactions only.  The PFSS model introduces nearest-neighbour
interactions additionally to the zero-range interactions of the
ZRP. We will use the simpler ZRP to setup and validate our methods to
estimate the time scales of the condensation process in the latter
model.

\begin{figure}
  \centering
  \includegraphics{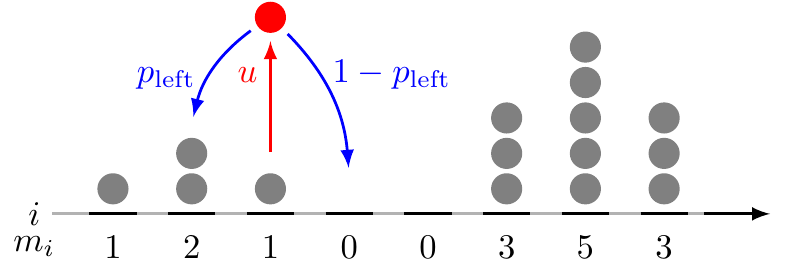}
  \caption{Schematics of the discussed models.  From a randomly
    selected site $i$ a particle is taken with probability
    $u_{\text{ZRP}}=u(m_{i})/u_{\text{max,ZRP}}$ or
    $u_{\text{PFSS}}=u(m_{i}\vert
    \revise{m_{i-1},m_{i+1}})/u_{\text{max,PFSS}}$, and then moved to
    sites $i-1$ or $i+1$ with probability $p_{\text{left}}$ or
    $1-p_{\text{left}}$, respectively.}
  \label{fig:zrp-scheme}
\end{figure}

In both models the system consists of a number of indistinguishable
particles that occupy the sites of a lattice or graph. The number of
particles occupying a specific site $i$ is called the occupation
number $m_{i}$ and can be any integer value $0\le m_{i}\le M$, where
$M=\sum_{i=1}^{L}m_{i} = \rho L$ is the total particle number in a
system of $L$ sites. At every step of the stochastic process a site
$i$ in the lattice is randomly selected for trial of particle hopping.
If the site is occupied, hopping of a single particle succeeds with
the hopping rate $u$ per unit time and the particle hops to a
neighbour of its departure site. Random choice of the hopping target
leads to an equilibrium system, while a biased choice can lead to a
system driven far from equilibrium. We will consider the case of the
one-dimensional periodic lattice, where such a drive is easily
implemented by a probability $p_{\text{left}}$ of the particle to hop
to the nearest neighbour in negative direction of the lattice and in
positive direction with probability $1-p_{\text{left}}$.  For any
given lattice and number of particles $M$ the individual dynamics of
the particles in these models are therefore fully described by the
hopping rate function and the strength of the bias in the hopping
direction.  The general update scheme of this family of processes is
given in Fig.~\ref{fig:zrp-scheme}.

\subsection{The zero-range process}
\label{sec:zrp-transport}

The ZRP was originally proposed to model a set of non-exclusively
interacting random walks by Spitzer~\cite{Spitzer1970}.  It is
constructed to be analytically tractable and has, for symmetric as
well as asymmetric dynamics, always the same factorised stationary
distribution
\begin{equation}
  \label{eq:steady-state-zrp} P_{\text{ZRP}}(\{m\}) =
\frac{1}{Z_{\text{ZRP}}^{M,L}} \left( \prod_{i=1}^{L} p(m_{i}) \right)
\delta_{\textstyle\sum_{i=1}^{L} m_{i}, M}\; ,
\end{equation} where $\{m\}=(m_{1}, \ldots, m_{L})$ is a specific
configuration of the system. The factorisation of the steady state is
carried out over the single-site weight functions $p(m)$ for all
sites.  The normalisation constant $Z_{\text{ZRP}}^{M,L}$ used here
has the same role as the partition function in an equilibrium model
and the Kronecker symbol $\delta_{\sum_{i=1}^{L} m_{i},M}$ ensures the
conservation of particles. With the choice of the weight function
\begin{equation}
  \label{eq:weights-zrp} p(m) = \prod_{n=1}^{m} \left( \frac{1}{u(n)}
\right),\; p(0)=1
\end{equation} the hopping rate function then becomes
\begin{equation}
  \label{eq:hopping-weights-zrp} u(m) = \frac{p(m-1)}{p(m)}.
\end{equation} It is easy to check that this leads to the steady state
(\ref{eq:steady-state-zrp}) by inserting (\ref{eq:weights-zrp}) and
(\ref{eq:hopping-weights-zrp}) into the master equation
\begin{align}
  \label{eq:zrp-master-equation} \frac{\partial}{\partial t}
P\left(\{m\}\right) = 0 = & \sum_{\{m\}'} \Big[ W\left(\{m\}' \to
\{m\}\right)P\left(\{m\}',t\right) \nonumber \\ & -
W\left(\{m\}\to\{m\}'\right)P\left(\{m\},t\right) \Big],
\end{align}
\revise{where the transition rates are
  $W\left(\{m\}\to\{m\}'\right) = u(m_i)p_{\text{left}}$ for a hop to
  the left
  ($\{m\}'=\{m_1, \ldots, m_{i-1}+1, m_{i}-1,\ldots, m_{L}\}$),
  $W\left(\{m\}\to\{m\}'\right) = u(m_i)(1-p_{\text{left}})$ for a hop
  to the right
  ($\{m\}'=\{m_1, \ldots, m_{i}-1, m_{i+1}+1,\ldots, m_{L}\}$), and
  $W\left(\{m\}\to\{m\}'\right) = 0$ otherwise.}

\begin{figure} \centering
  \includegraphics{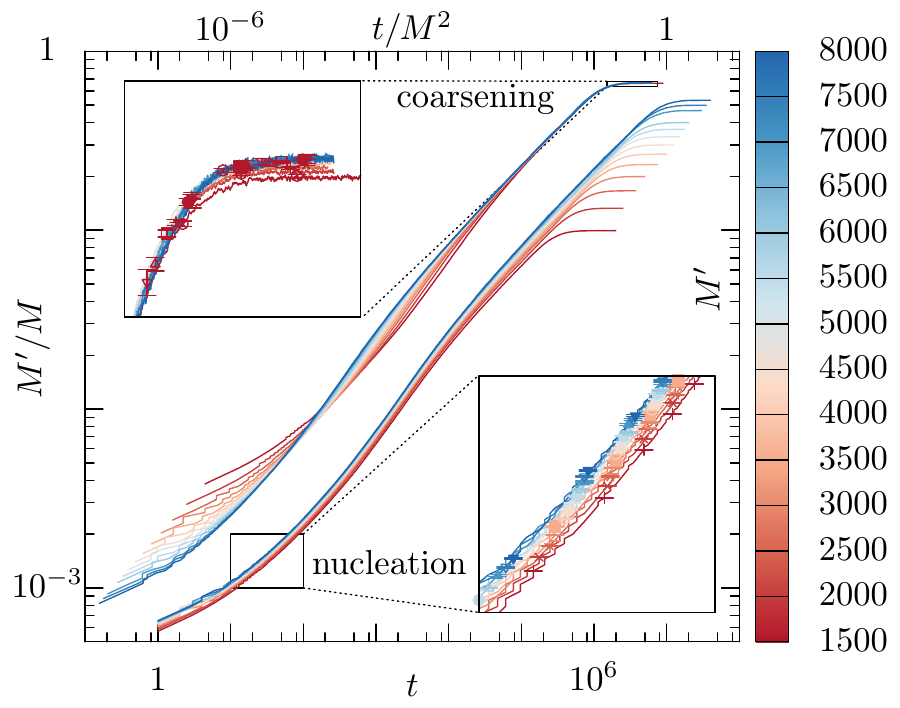}
  \caption{The nucleation (bottom right) and coarsening (top left)
    processes for various system sizes in the ZRP as observed in the
    average mass of the largest droplet at time $t$. The two groups of
    curves show the same data. The upper group, plotted versus the top
    and left axes, is merely rescaled in time by $1/M^{2}$ and in mass
    by $1/M$ in order to collapse the trajectories of different system
    sizes to a single master curve \revise{for late times}, as shown
    in the upper inset. The lower group, plotted versus the lower and
    right axes, is shown without rescaling.  The lower inset shows the
    weak dependence of droplet growth on system size in the early
    condensation dynamics.  The curves were determined for totally
    asymmetric dynamics $p_{\text{left}}=1$ using 1000 trajectories
    per size $L=M$.}
  \label{fig:nucleation-coarsening}
\end{figure}

In this work we consider the \revise{well-known} ZRP condensation
model~\revise{\cite{OLoan1998,Evans2000}} with hopping rate function
\begin{equation}
  \label{eq:hopping-zrp} u(m) = 1 + \frac{b}{m}.
\end{equation}
For this choice of dynamics, condensation occurs for \hbox{$b>2$}
above a critical density of $\rho_{\text{c}}=1/(b-2)$. When the total
number of particles is increased above the condensation threshold
$M>\rho_{\text{c}}L$, the translational symmetry breaks down and the
excess particles $M'=M-\rho_{\text{c}}L$ condense at a single site in
the system. The occupation number of the rest of the system remains on
average at the critical density $\rho_{\text{c}}$.  For illustration,
the average mass of the largest droplet $M'(t)$ during a complete
condensation is shown in Fig.~\ref{fig:nucleation-coarsening} for a
number of system sizes with totally asymmetric dynamics. Note, how for
small times, the largest droplets grow very similarly, while
coarsening takes much longer for larger lattices.

The scaling of the nucleation and coarsening times as determined by
Grosskinsky et al.~\cite{Grosskinsky2003} assumes a power-law in the
system size of the form
\begin{equation} \tau = a M^{\delta},
  \label{eq:scaling}
\end{equation} that is also observed for other relaxation processes in
the ZRP with different dynamics~\cite{Barma2002}.  To estimate the
nucleation time scale, Grosskinsky et al.\ propose a mass threshold
$m_{\text{t}}$ for a droplet nuclei and calculate the time it takes to
populate the system with such nuclei.  For the choice of a threshold
linear in the system size
$m_{\text{t,lin}}=\alpha_{\text{lin}}(\rho-\rho_{\text{c}})L$, with some
constant $\alpha \ll 1$, they estimate the scaling exponent of the
nucleation time as $\delta_{\text{nucl,ZRP}}=3$ for symmetric
($p_{\text{left}}=1/2$) and $\delta_{\text{nucl,ZRP}}=2$ for totally
asymmetric ($p_{\text{left}}=1$) dynamics.  The coarsening time is
dominated by the typical time for a droplet to loose all particles,
which is determined using the fact that for the ZRP the evaporation
rate of a droplet is simply the hopping rate for the droplet mass.

\subsection{The PFSS transport process}
\label{sec:pfss-transport}

\begin{figure*}
  \includegraphics{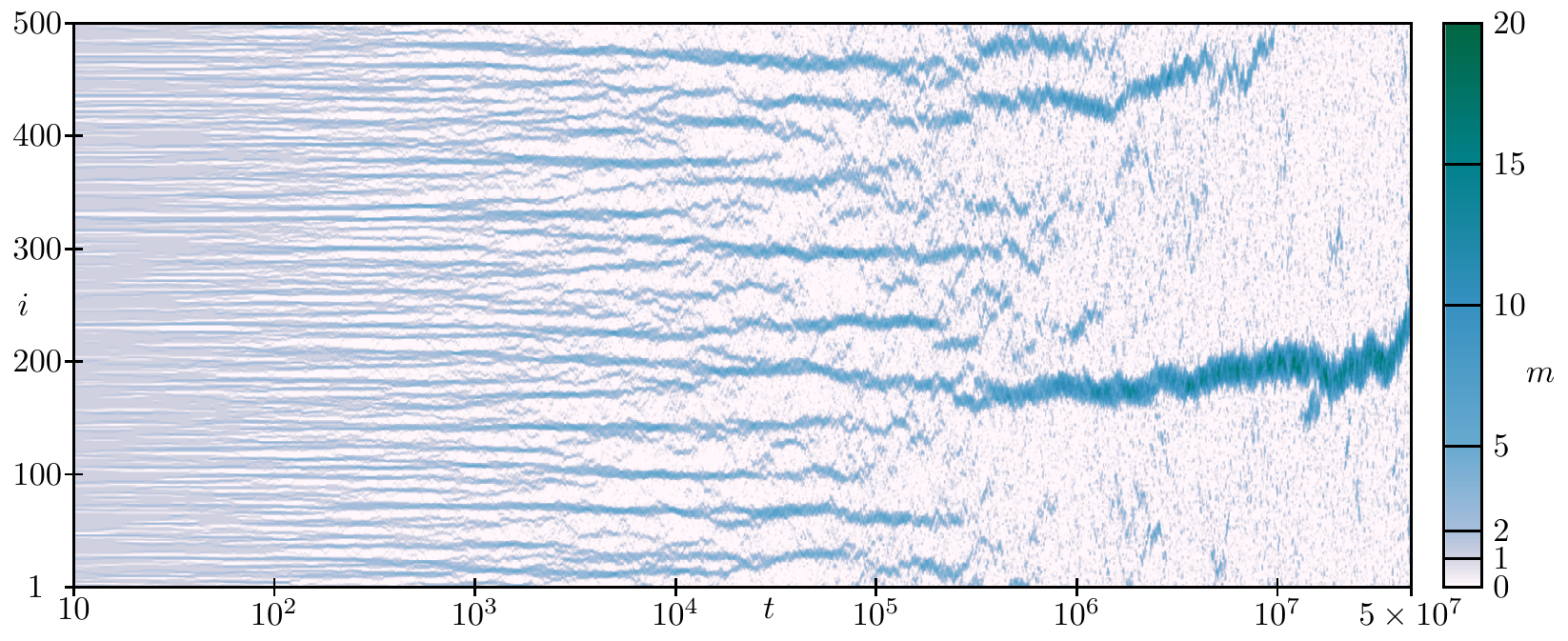}
  \caption{Sample time series of a PFSS condensation process in a
    system of $L=500$ sites and $M=500$ particles for totally
    asymmetric dynamics ($p_{\text{left}}=1$, ``particles cyclically
    move down'') and interaction parameters $U=J=1$.  Each vertical
    slice in the plot corresponds to the occupation number vector at
    time $t$. The time axis is logarithmic to show the different
    stages of the condensation process.  Starting with a disordered
    state in the leftmost slice, small droplets emerge and gradually
    coarsen on every time scale until a single condensate remains at
    about $t\approx 10^{7}$.}
  \label{fig:condensation-sample}
\end{figure*}

The basic PFSS short-range interaction transport process differs from
the ZRP by taking the interactions between particles on adjacent sites
into account.  This is realized by replacing the single-site weight
$p(m)$ of the ZRP with a two-point weight function $g(m,n)$ for each
bond in the system. The stationary distribution of this process then
assumes the form
\begin{equation}
  \label{eq:steady-state-pfss} P(\{m\}) =
\frac{1}{Z_{\text{PFSS}}^{M,L}} \prod_{i=1}^{L} g(m_{i},m_{i+1})
\delta_{\textstyle\sum_{i=1}^{L} m_{i},M},
\end{equation} i.e.\ it factorizes over pairs of sites, giving rise to
the name \emph{pair-factorized steady states (PFSS)} transport process.
Here the bond weight function $g(m,n)$ is a positive and in our case
symmetric function. The corresponding hopping rate function that
controls the dynamics of the process is determined similarly as for
the ZRP \revise{using the master
  equation~\eqref{eq:zrp-master-equation} with appropriate transition
  rates
  $W\left(\{m\}\to\{m\}'\right) = u(m_i\vert m_{i-1},
  m_{i+1})p_{\text{left}}$ for a hop to the left
  ($\{m\}'=\{m_1, \ldots, m_{i-1}+1, m_{i}-1,\ldots, m_{L}\}$),
  $W\left(\{m\}\to\{m\}'\right) = u(m_i\vert m_{i-1},
  m_{i+1})(1-p_{\text{left}})$ for a hop to the right
  ($\{m\}'=\{m_1, \ldots, m_{i}-1, m_{i+1}+1,\ldots, m_{L}\}$), and
  $W\left(\{m\}\to\{m\}'\right) = 0$ otherwise.
The resulting hopping rate function is then given by}
\begin{equation}
  \label{eq:hopping-rate-pfss} u(m_{i}\vert m_{i-1}, m_{i+1})=
\frac{g(m_{i}-1,m_{i-1})}{g(m_{i},m_{i-1})}
\frac{g(m_{i}-1,m_{i+1})}{g(m_{i},m_{i+1})}.
\end{equation}

A compelling general choice of the bond weight is to use the
factorized form $g(m,n)=\sqrt{p(m)p(n)}K(\left\vert m-n \right\vert)$,
with a zero-range site weight function $p(m)$ as in the ZRP and a
short-range part $K(x)$.  Thus, the ZRP is easily reproduced as a
special case by setting $K(x) \equiv 1$ and a broad range of
interactions can be implemented by interpreting the partial weights
$p(m)$ as a particle-site potential and $K(\vert m - n \vert)$ as a
particle-particle interaction term.  Due to the ranged interactions
there is now the possibility that spatially extended condensates
emerge in the system. A number of specific weights and thus dynamics
have been considered in the literature for this model, such as that
suggested in its original proposition by Evans et~al.~\cite{Evans2006}
or the tunable interactions resulting in various condensate shapes
\cite{Waclaw2009c,Waclaw2009b,Ehrenpreis2014}.

The main question that we are going to resolve in this work, is
whether the nucleation and coarsening dynamics of condensation in this
model with short-range interactions takes place on the same time scale
as for the ZRP. To do this, we will consider the dynamics suggested in
Ref.~\cite{Evans2006} given by the partial weights
\begin{equation}
  \label{eq:pfss-evans-weights} p(m)=\mathrm{e}^{U \delta_{m,0}},
\quad K(x)=\mathrm{e}^{-Jx}.
\end{equation} Here, $U$ gives the strength of an on-site potential
and $J$ can be interpreted as a surface energy. In this work, we will
only explicitly discuss the case $U=1, J=1$, although we did check
different parametrisations as well as different dynamics with
pair-factorized steady states featuring condensation.  Inserting this
into the Eq.~\eqref{eq:hopping-rate-pfss} gives the hopping rate
function
\begin{equation}
  \label{eq:pfss-evans-hopping} u(m_{i} \vert m_{i-1}, m_{i+1}) =
  \begin{cases}
    \mathrm{e}^{-2J+U\delta_{m_{i},1}} & m_{i} \le m_{i-1}, m_{i+1}, \\
    \mathrm{e}^{2J + U\delta_{m_{i},1}} & m_{i} > m_{i-1}, m_{i+1}, \\
    \mathrm{e}^{U\delta_{m,1}} & \text{otherwise.}
  \end{cases}
\end{equation} As an illustration of this dynamics,
Fig.~\ref{fig:condensation-sample} shows a time series of
configurations from a random initial state to the stationary state
with a single condensate.

For this choice of dynamics the critical density can be determined as
$\rho_{\text{c}}=0.2397$ for parameters $U=1$, $J=1$~\cite{Waclaw2009c}.
Above this density an extended condensate of roughly parabolic shape
containing the excess particles $(\rho -\rho_{\text{c}})L$ emerges.
The exact shape can be determined as well~\cite{Waclaw2009b}.  The
extension of the condensate scales as the square root of its mass $M'$
for sufficiently large systems~\cite{Evans2006,Waclaw2009c}. Because
of the spatial extension as well as the smooth shape of the droplets,
and ultimately the remaining single condensate, it is non-trivial to
reproduce either the analytical or the qualitative arguments for the
nucleation and coarsening time scales in this case.  Therefore, in
this work, we perform numerical simulations of both the ZRP and PFSS
model to calibrate our methods by the ZRP results and determine the
time-scale exponents for the PFSS process using the power-law ansatz
\eqref{eq:scaling} as for the ZRP.

\section{Numerical simulations}
\label{sec:methods}

We employ a kinetic Monte Carlo method with discrete time steps that
directly simulates the dynamics of the considered transport
processes. Per time step, every lattice site is considered on average
once for a local update that consists of the following steps: A site
$i$ with $1 \le i \le L$ is randomly and uniformly chosen and, if
occupied, the hopping rate $u_{i}$ for that site is determined
according to~\eqref{eq:hopping-zrp} or~\eqref{eq:pfss-evans-hopping}.
A particle leaves the site, with the probability
$u_{i}/u_{\text{max}}$, where
\begin{equation} u_{\text{max}} =
  \begin{cases}
    b+1 & \text{ZRP} \\
    \mathrm{e}^{2J+U} & \text{PFSS for } U,J > 0
  \end{cases}
\end{equation}
is the maximal hopping rate for the given model. The particle then
enters the left ($i-1$) or right ($i+1$) \revise{neighbour} with
probability $p_{\text{left}}$ and $1-p_{\text{left}}$, respectively.
At the boundary sites $i=1,L$ we consider periodic boundary
conditions, so that the left neighbour of the first site and the right
neighbour of the last site are identified as the last and first site
respectively. Because of the definition of the hopping rate in events
per physical time unit, a Monte Carlo \emph{sweep} of $L$ single-site
updates corresponds to $1/u_{\text{max}}$ units of physical time. By
repeating this update procedure, a single trajectory of configurations
$\{m\}(t)$ is generated.

In contrast to steady-state simulations, where an observable can be
estimated from a large number of observations in a single trajectory,
here every trajectory contributes only a single observation to the
estimation of the typical nucleation and coarsening times. To
determine a sufficiently good estimate for the scaling exponents we
hence generated on the order of several thousand trajectories for
every parametrisation.

\section{Nucleation}
\label{sec:nucleation}

Nucleation is the first stage of the condensation process, covering
the formation of particle droplet nuclei from the bulk system until
the bulk phase is relaxed and coarsening takes over. Since the number
of particles in the system $M=\rho L$ is conserved, the initial
configuration is prepared using a uniform distribution of
particles. For any supercritical density $\rho>\rho_{\text{c}}$ the
system then contains $(\rho-\rho_{\text{c}})L$ excess particles that
settle at random sites forming droplets. In the case of the ZRP, such
a droplet is just a larger number of particles on a single site with
respect to the occupation number distribution in the bulk of the
system. The mass of a droplet, that is the number of contained
particles, is therefore equal to the occupation number at its location
for the ZRP.  In the PFSS model, where droplets are in general
extended, we require separation by at least one site with occupation
below the critical density $m_{i}<\rho_{\text{c}}$, that is $m_{i}=0$
in both models for the parameters considered in this paper. The sum of
occupation numbers in the droplet region then defines its mass.

To gain a basic understanding of the involved time scales determining
the nucleation process, we can somewhat generalize the heuristic
approximation of the scaling exponent for the ZRP by Grosskinsky et
al.~\cite{Grosskinsky2003}. When the typical cluster mass is assumed
to be linear in the system size $\alpha (\rho-\rho_{\text{c}})L$, it
follows that there are on the order of $1/\alpha$ droplets in the
system. The density of droplets in turn determines the average free
distance that particles travel to a nuclei. The time, in units of
single hops, for on the order of $L$ particles to travel a distance of
order $L$ is then roughly $L^{2}$ for totally asymmetric dynamics and
$L^{3}$ including diffusive motion due to symmetric dynamics.  The
choice of the typical droplet mass being linear in the system size
ensures that droplets are distinctly larger than random fluctuations,
which are of order $\revise{[}\Gamma(b-1)L\revise{]}^{1/(b-1)}$ for
the ZRP~\cite{Grosskinsky2003}, where $\Gamma(x)$ is the gamma
function. However, assuming a different functional dependence of the
typical droplet mass, such as a proportionality to the square root of
the excess particles
$\alpha_{\text{sqrt}}\sqrt{(\rho-\rho_{\text{c}})L}$, results in a
change of the time scale as well: On the order of $L^{1/2}$ particles
travelling for a mean free distance of order $L^{1/2}$ sites yields a
time scale of order $L$ for asymmetric and $L^{3/2}$ for symmetric
dynamics.

Because of this, we will first reproduce and compare the time scale
for nucleation in the ZRP before proceeding to the PFSS
model. Furthermore, we will consider three types of typical droplet
masses of the form
\begin{equation}
  \label{eq:typical-masses} m_{\text{t}} =
  \begin{cases} \alpha_{\text{lin}}(\rho-\rho_{\text{c}})L &
m_{\text{t,lin}}\\ \alpha_{\text{sqrt}}\sqrt{(\rho-\rho_{\text{c}})L}
& m_{\text{t,sqrt}}\\ \text{const} & m_{\text{t,const}}
  \end{cases}
\end{equation} to check our former assumptions and compare the
resulting kinetics. The additional threshold $m_{\text{t,const}}$
independent of the system size is motivated as follows. The critical
droplet size in the nucleation regime is the size of a droplet for
which its growth rate becomes faster than its decay rate. For the
single-site droplets in the ZRP, we can determine this size by
evaluating the droplet growth in the vicinity of the critical droplet
size $m_{\text{c}}$
\begin{equation}
  \label{eq:growth-rates} \left.\frac{\partial m}{\partial
t}\right\vert_{m\approx m_{c}} = 0 = \sum_{n=1}^{\infty}P(m_{\pm
1}=n)u(n) - u(m_{c}),
\end{equation} where $P(m_{\pm 1}=n)$ is the probability to find $n$
particles on the adjacent sites. As the bulk should relax within the
nucleation time, we can assume that $P(m_{\pm 1}=n)=\pi(m)$ is the
mass distribution of the ZRP, that is $\pi(m) = p(m)\exp(-\mu m)$ with
the chemical potential $\mu$ in the grand-canonical
system
and thus $\pi(m) \propto p(m) \approx \Gamma(b+1)m!/\Gamma(b+1+m)$ for a
fixed number of particles. We can now find the critical value as
$m_{\text{c}}\approx b\Gamma(b-1) / \revise{[}\Gamma(b-1)+\Gamma(b-2)\revise{]}$ which is
$m_{\text{c}}=15/2$ for the interaction strength $b=5$ used here and
set our threshold including a small safety margin to
$m_{\text{t,const}}=10$.

To determine the nucleation time scale using numerical simulations, we
will compute the transition time when the coarsening of droplets
starts to dominate the kinetics of the system, that is the formation
of new droplets of at least the typical mass $m_{\text{t}}$ becomes
smaller than the evaporation of existing ones.  This is achieved by
counting the number of droplets that have at least the typical mass
(threshold mass) at every time in the trajectory and computing its
average over many trajectories.  The scaling exponent is then
determined visually for the best data collapse assuming the power
law~\eqref{eq:scaling}.

\subsection{ZRP}
\label{sec:nucleation-zrp}

\begin{figure*} \centering
  (a)\hspace{-1.5em}\includegraphics[width=0.33\textwidth]{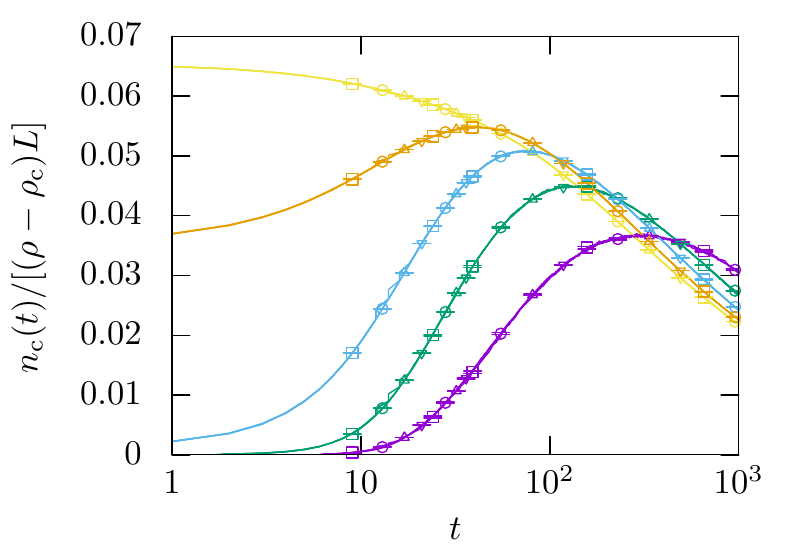}\hfill
  (b)\hspace{-1.5em}\includegraphics[width=0.33\textwidth]{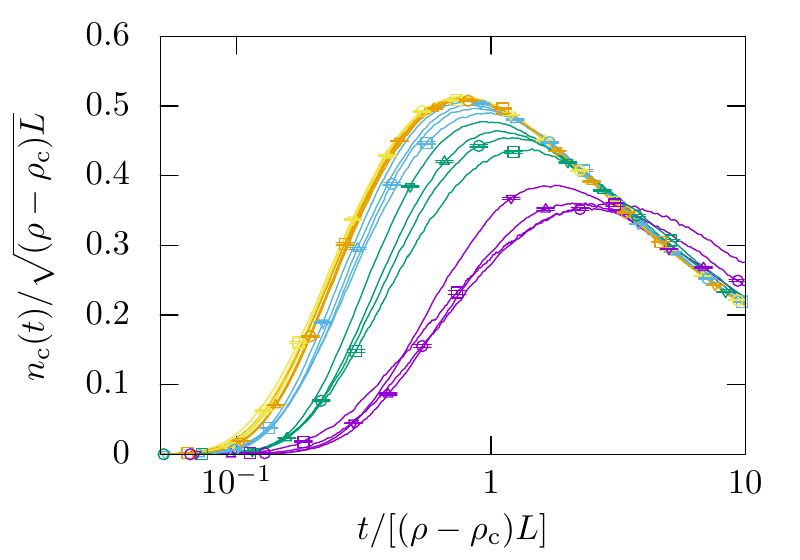}\hfill
  (c)\hspace{-1.5em}\includegraphics[width=0.33\textwidth]{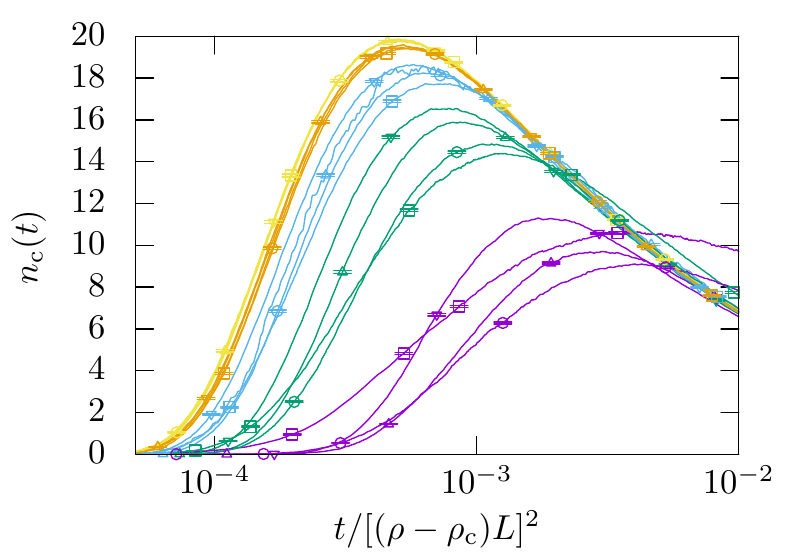}
  (d)\hspace{-1.5em}\includegraphics[width=0.33\textwidth]{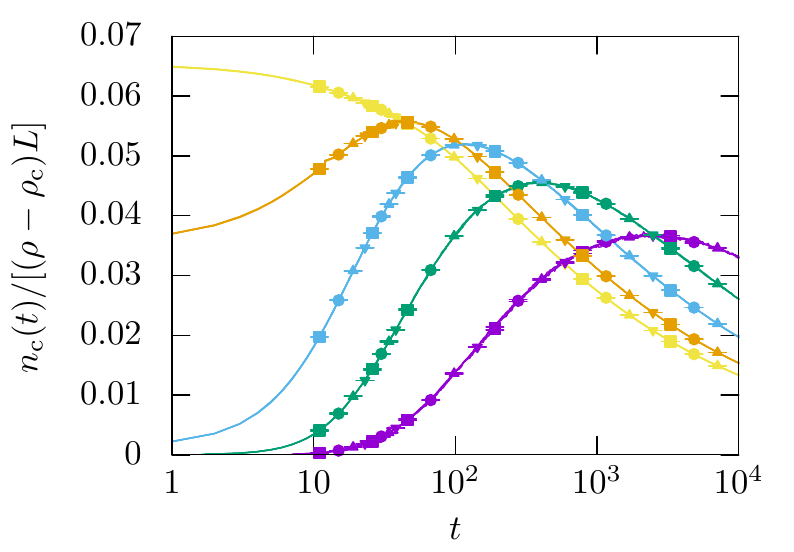}\hfill
  (e)\hspace{-1.5em}\includegraphics[width=0.33\textwidth]{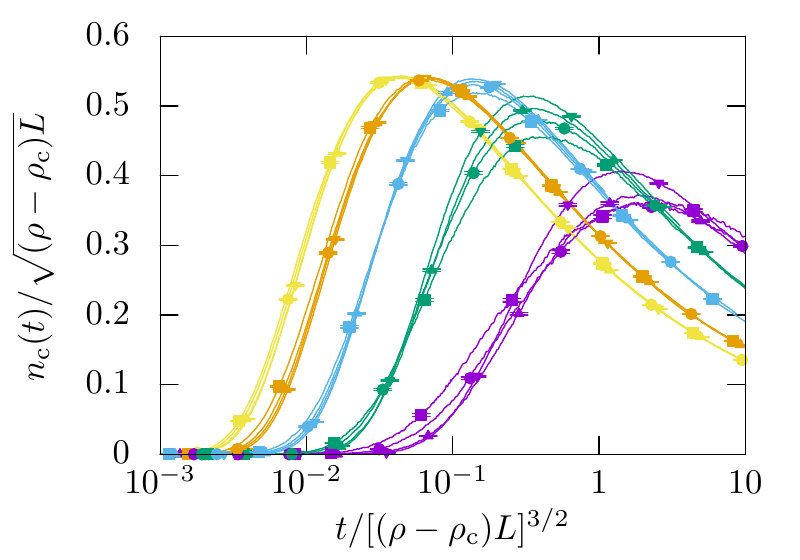}\hfill
  (f)\hspace{-1.5em}\includegraphics[width=0.33\textwidth]{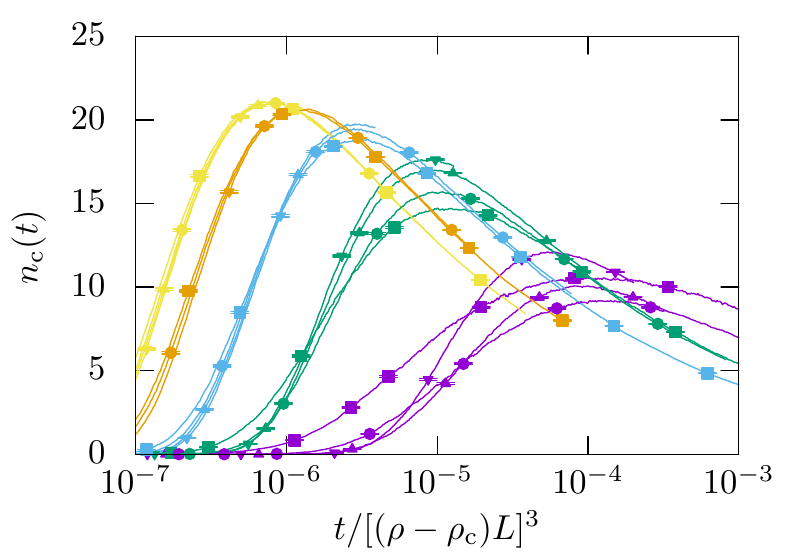}
  \includegraphics{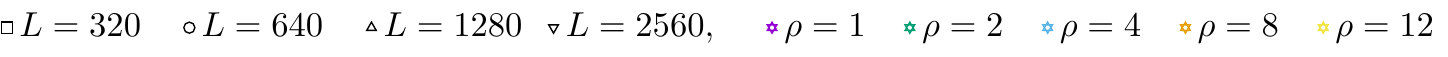}
  \caption{Droplet counts $n_{\text{c}}(t)$ for the ZRP with $b=5$ in
    the nucleation regime for thresholds in the droplet mass that are
    constant \revise{[$m_{\text{t,const}}=10$ in (a), (d)]}, grow as
    the square root
    \revise{[$m_{\text{t,sqrt}}=\sqrt{(\rho-\rho_{\text{c}})L}$ in
      (b), (e)]} of, or linearly
    \revise{[$m_{\text{t,lin}}=(1/40)(\rho-\rho_{\text{c}})L$ in (c),
      (f)]} to the total number of particles.  Results for totally
    asymmetric dynamics \revise{($p_{\text{left}}=1$)} are shown
    with empty symbols in the top row, those for symmetric dynamics
    \revise{($p_{\text{left}}=1/2$)} with filled symbols in the bottom
    row. Symbols represent the different system sizes
    $L=320, 640, 1280$, and $2560$, colours indicate different
    particle densities $\rho=1, 2, 4, 8$, and $12$. For readability
    symbols are only sparsely plotted and mark the corresponding lines
    that fully reflect the available data. The resulting time scales
    on the x-axis roughly correspond to the heuristic approximation.}
  \label{fig:nucleation-zrp}
\end{figure*}

To compare with the results of Grosskinsky
et~al.~\cite{Grosskinsky2003} and validate our numerical methods, we
use a similar parameter set with system sizes of $L=320, 640, 1280$,
and $2560$ sites, interaction strength $b=5$, yielding
$\rho_{\text{c}} = 1/(b-2) = 1/3$, and somewhat more variation in the
particle density $\rho=1, 2, 4, 8$, and $12$.  To compute the average
droplet-count function $n_{\text{c}}(t)$ we simulated 1200
trajectories for each combination of parameters with totally
asymmetric ($p_{\text{left}}=1$) as well as symmetric
($p_{\text{left}}=1/2$) dynamics. We then rescaled the average droplet
counts to achieve a good collapse for different system sizes and get
the scaling with respect to the mass
thresholds~\eqref{eq:typical-masses}.

Figure~\ref{fig:nucleation-zrp} shows that these results indeed
reflect the heuristic approximations made for the different
assumptions of the typical droplet mass $m_{\text{t}}$. The curves of
the droplet counts collapse in two ways. Firstly, for identical values
of the particle density $\rho$ there is a collapse for individual
system sizes $L$.  Secondly, for the size-dependent thresholds and
asymmetric dynamics, this collapse increases with the particle density
and a common master curve is approached. For typical droplet masses
$m_{\text{t,const}}$ independent of the system size, the collapse for
different $L$ within the same density is remarkable.  While this might
to some extent reflect the higher absolute value of droplet counts and
thus greater amount of self-averaging in that situation, it also
suggests that this assumption is indeed physical.
The rescaling factors in the vertical axis for the mass thresholds
$m_{\text{t,const}}$ and $m_{\text{t,sqrt}}$ reflect the respective
expected number of droplets per system size according to the heuristic
approximation.
Comparing with Ref.~\cite{Grosskinsky2003}, that is specifically for
typical droplet sizes
$m_{\text{t,lin}}=\alpha_{\text{lin}}(\rho-\rho_{\text{c}})L$ linear
in the system size with $\alpha_{\text{lin}}=1/40$, our value of the
scaling exponent $\delta_{\text{nucl}}=2$ is consistent with the value
determined for the assumption therein.

While the droplet mass threshold in the original work by Grosskinsky
et al.~\cite{Grosskinsky2003} is chosen to ensure that such droplets
are distinct from fluctuations of order
$\revise{[}\Gamma(b-1)L\revise{]}^{1/(b-1)}$ in the initial system for
any value of the interaction strength $b$, our assumption
$m_{\text{t,sqrt}}$ fulfills this requirement as well for \revise{the}
considered interaction strength of $b=5$. Additionally, we performed a
similar amount of simulations for $b=4$ yielding very similar
results. Considering that the data collapse for $m_{\text{t,sqrt}}$
and $m_{\text{t,lin}}$ is equally good, we would then suggest
that\revise{,} if the duration of the nucleation regime can be limited
in this approach, the systematically smaller yet valid assumption of
\revise{the} typical droplet size should be physical.

\begin{figure*} \centering
  (a)\hspace{-1.5em}\includegraphics[width=0.33\textwidth]{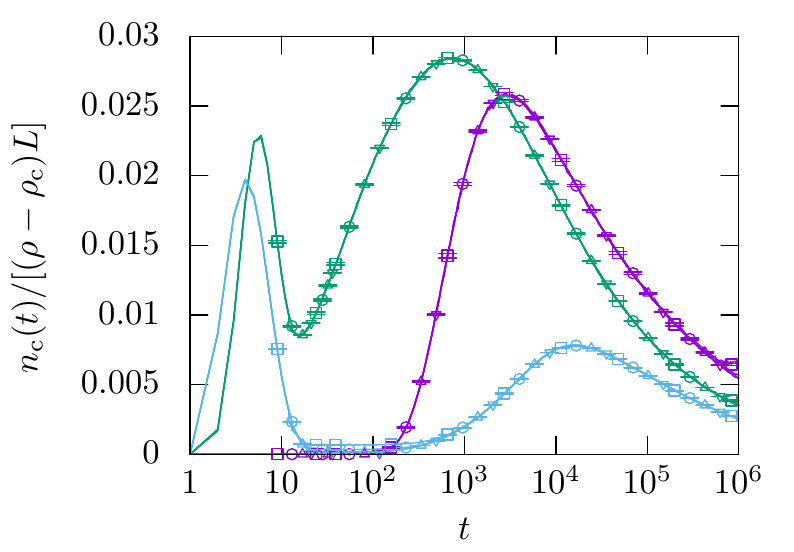}\hfill 
  (b)\hspace{-1.5em}\includegraphics[width=0.33\textwidth]{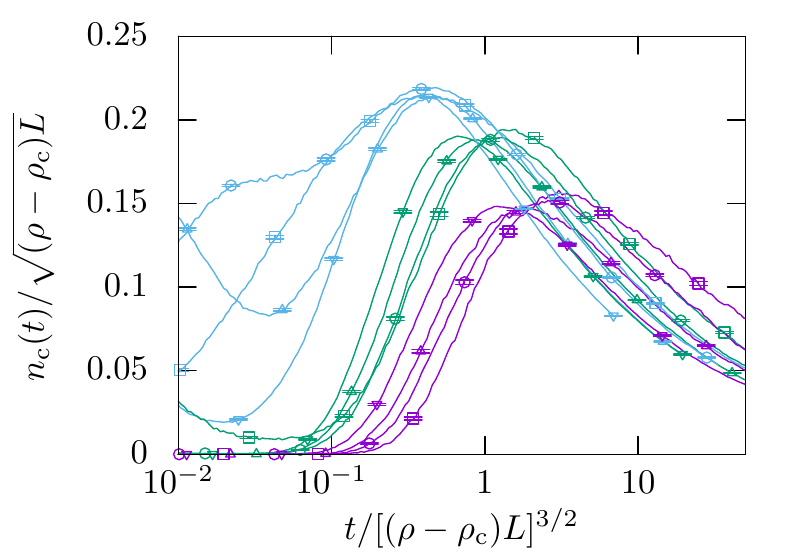}\hfill 
  (c)\hspace{-1.5em}\includegraphics[width=0.33\textwidth]{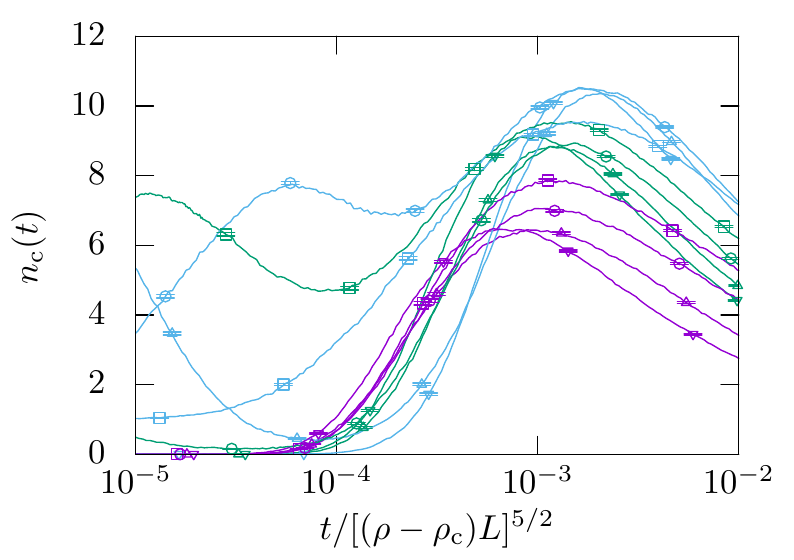} 
  (d)\hspace{-1.5em}\includegraphics[width=0.33\textwidth]{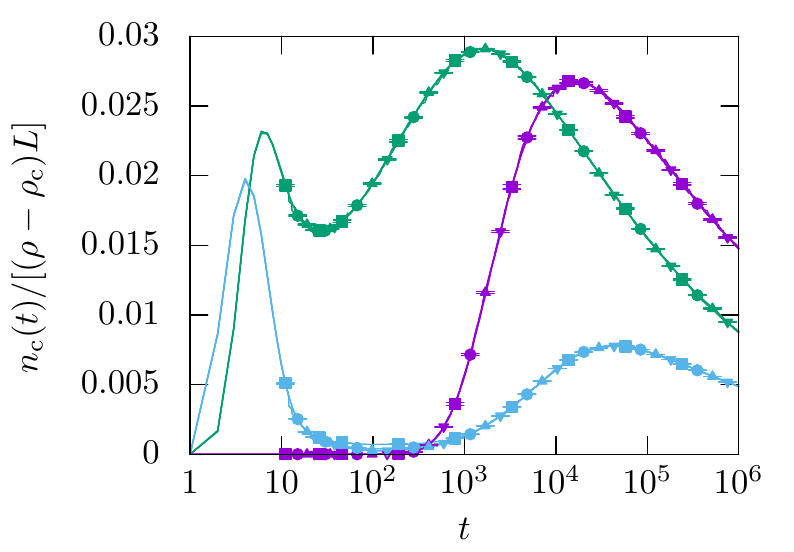}\hfill 
  (e)\hspace{-1.5em}\includegraphics[width=0.33\textwidth]{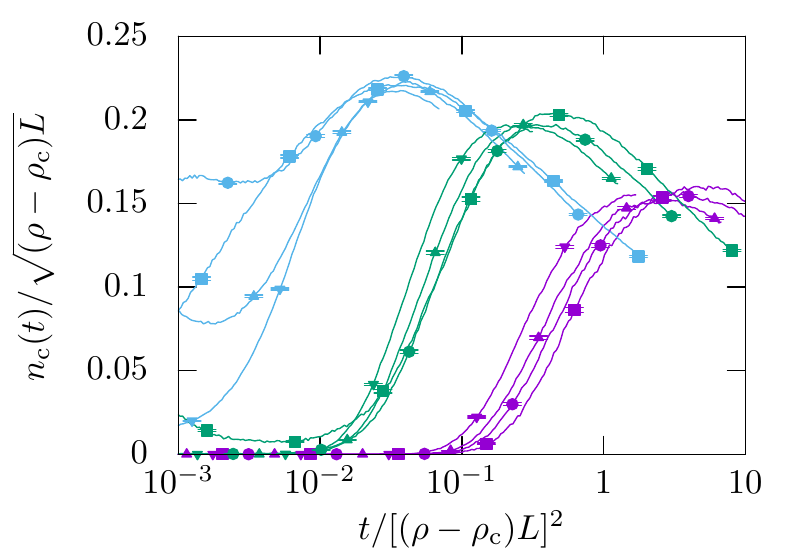}\hfill 
  (f)\hspace{-1.5em}\includegraphics[width=0.33\textwidth]{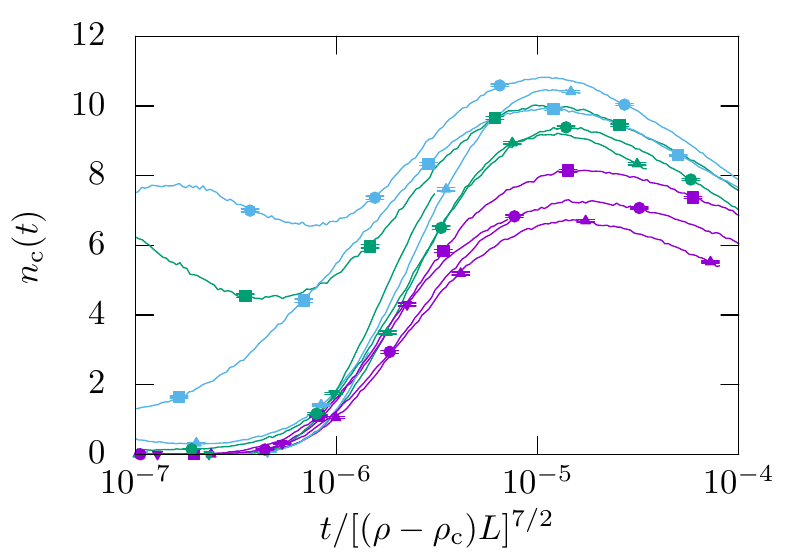}
  \includegraphics{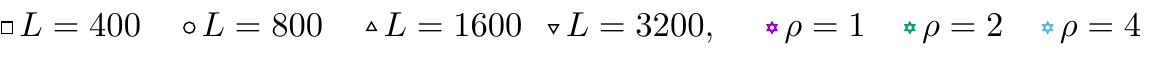}
  \caption{Droplet counts $n_{\text{c}}(t)$for the PFSS with $U=J=1$
    in the nucleation regime for thresholds in the droplet mass that
    are constant \revise{[$m_{\text{t,const}}=20$ in (a), (d)]}, grow
    as the square root
    \revise{[$m_{\text{t,sqrt}}=(5/2)\sqrt{(\rho-\rho_{\text{c}})L}$
      in (b), (e)]} of, or linearly
    \revise{[$m_{\text{t,lin}}=(1/20)(\rho-\rho_{\text{c}})L$ in (c),
      (f)]} to the total number of particles.  Results for totally
    asymmetric dynamics \revise{($p_{\text{left}}=1$)} are shown with
    empty symbols in the top row, those for symmetric dynamics
    \revise{($p_{\text{left}}=1/2$)} with filled symbols in the bottom
    row. Symbols represent the different system sizes
    $L=400, 800, 1600$, and $3200$, colours indicate different
    particle densities $\rho=1, 2$, and $4$.}
  \label{fig:nucleation-pfss}
\end{figure*}

Finally, as the third assumption for the typical droplet size
$m_{\text{t,const}}$, we use our estimate of the critical droplet size
for the ZRP $m_{\text{t,const}}=10$ although it becomes smaller than
the expected fluctuations in the initial configuration for larger
systems.  This is visible in
Figs.~\ref{fig:nucleation-zrp}(a,d)\revise{,} where for large
densities the droplet count indeed approaches from higher values and
does not reach a local maximum.  Using a threshold with a larger
margin to the critical droplet size, such as $m_{\text{t,const}}=20$,
mitigates this effect so that the droplet counts for large densities
become similar to those for lower densities $\rho\le 8$.  Without any
rescaling in time, the data collapse in the observed droplet counts
for systems of different sizes is excellent within the same
particle density.

Our results show that, in order to determine a meaningful time scale
for the nucleation process using this method, it is crucial to
carefully select a relation of the typical droplet mass $m_{\text{t}}$
to the system size with respect to the considered system and to some
extent the desired properties of the forming droplets. With respect to
that, we only visually determined estimates of the nucleation
exponents instead of performing extensive analysis to find numerical
values for maximum overlap.  A summary of the determined scaling
exponents is given in Table~\ref{tab:nucleation-exponents}.
Furthermore, we validated our numerical methods to proceed to the
nucleation time of the PFSS model in the next section.

\subsection{PFSS}
\label{sec:nucleation-pfss}

For the PFSS we use a fixed set of interaction parameters $U=J=1$ for
the weights~\eqref{eq:pfss-evans-weights} and resulting hopping rate
function~\eqref{eq:pfss-evans-hopping}.  The remaining parameters are
comparable to those for the ZRP with a range of system sizes of
$L=400, 800, 1600$, and $3200$ sites and particle densities of
$\rho=1, 2$, and $4$. We use the same method as for the ZRP to
determine the scaling exponents $\delta_{\text{nucl}}$ for the typical
droplet sizes~\eqref{eq:typical-masses} with slightly larger constants
$\alpha_{\text{lin}}=1/20$ and $\alpha_{\text{sqrt}}=5/2$ due to the
lower critical density $\rho_{\text{c}}\approx0.2397$ and the spatial
extension of the droplets. As for the ZRP, we also consider a typical
droplet mass, that we merely choose somewhat larger as
$m_{\text{t,const}}=20$ instead of estimating the critical droplet
size using the bulk particle distribution given in
Ref.~\cite{Waclaw2009b} and taking the much more involved particle
exchange at the droplet boundaries into account.

Due to the need of a separating site between individual droplets to
distinctly count them, the initial preparation must be carried out
with care, as a uniform distribution of particles easily leads to the
identification of a number of large droplets in the initial
configuration. This effect is visible as a small peak before the
larger droplet-count maximum in
Fig.~\ref{fig:nucleation-pfss}(a,d). To eliminate this influence on
the scaling exponents, we cross-checked our results using different
preparation schemes. While there is an influence on the form of the
droplet-count function $n_{\text{c}}(t)$, the effect on the rescaling
exponent is negligible.

Figure~\ref{fig:nucleation-pfss} shows rescaled droplet counts
determined for totally asymmetric as well as symmetric dynamics and
averaged from 1200 individual trajectories. Except for the constant
droplet-mass threshold, the obtained data collapse for the system
sizes that were feasible to simulate is less convincing than for the
ZRP. Due to the extension of droplets, the finite-size effects are
much stronger for the PFSS.

The observed scaling exponents with thresholds $m_{\text{t,lin}}$ and
$m_{\text{t,sqrt}}$ for the PFSS model, as given in
Table~\ref{tab:nucleation-exponents}, are larger by $1/2$ with respect
to those determined for the ZRP. From the heuristic approximation, one
could arguably expect a small decrease in the exponents, as the
distance particles need to travel to a nuclei is reduced by the
droplets' spatial extension. For the constant threshold droplet mass
$m_{\text{t,const}}$ we again observe a nearly perfect data collapse
without any rescaling in time at all as for the ZRP. We thus obtain
$\delta_{\text{nucl}}=0$ as the nucleation exponent independent of the
strength of the external drive $p_{\text{left}}$.

\begin{table}
  \caption{Estimated values of the scaling exponents $\delta_{\text{nucl}}$
    of the typical nucleation
    time for the ZRP and PFSS models with totally asymmetric and symmetric
    dynamics, respectively. }
  \label{tab:nucleation-exponents}
  \begin{tabular*}{\columnwidth}{l@{\extracolsep{\stretch{1}}}*{3}c}
    \hline\hline
    & {$m_{\text{t,const}}$} 
    & {$m_{\text{t,sqrt}}$}
    & {$m_{\text{t,lin}}$} \\
    \hline
    \;\,ZRP $\delta_{\text{nucl}}$ for $p_{\text{left}} = 1$ \hspace{1em} & $0$ & $1$ & $2$ \\
    \;\,ZRP $\delta_{\text{nucl}}$ for $p_{\text{left}} = 1/2$ & $0$ & $3/2$ & $3$ \\
    PFSS $\delta_{\text{nucl}}$ for $p_{\text{left}} = 1$  & $0$ & $3/2$ & $5/2$ \\
    PFSS $\delta_{\text{nucl}}$ for $p_{\text{left}} = 1/2$ & $0$ & $2$ & $7/2$ \\
    \hline\hline
  \end{tabular*}
\end{table}

Again, as for the ZRP, we would like to emphasize that this variety of
time scales depends on the choice of the typical droplet mass. For a
meaningful time scale, this threshold must be selected carefully.


\section{Coarsening}
\label{sec:coarsening}

After the relaxation of the bulk system and formation of droplet
nuclei the coarsening stage dominates the kinetics. The droplets in
the system exchange particles through the bulk.  Because larger
droplets emit less particles than smaller ones, the former grow while
the latter slowly evaporate and eventually disappear. As there are
fewer remaining droplets, the difference of the evaporation rates
becomes smaller compared to the droplet mass and the distance of
particle exchange grows. For the ZRP, the droplet evaporation rate is
directly given by the hopping rate function so that the coarsening
time scale can be determined quantitatively as $\delta=2$ for totally
asymmetric and $\delta=3$ for symmetric
dynamics~\cite{Grosskinsky2003}.

We will consider the case of the ZRP for totally asymmetric and
symmetric dynamics only to validate our methods and additionally
determine the scaling exponents for the coarsening time in the
situation of partially asymmetric dynamics for
$0.5\le p_{\text{left}} \le 1$. We will then proceed to determine the
scaling exponents for the expected coarsening time in the PFSS model
for asymmetric and symmetric dynamics $0.5\le p_{\text{left}} \le 1$.

In the remaining section, we will first describe the considered
methods to determine these scaling exponents and then discuss the
respective results for the ZRP as well as the PFSS.

A simple method to estimate the coarsening time for a single
trajectory is to compare the mass of the largest condensate with a
threshold at the expected value of the excess mass $M'$. The time of
first passage above this threshold from an initial state is recorded
as one measurement of condensation time. The threshold value must be
chosen slightly below the expected mass of the final condensate $M'$,
so that \revise{the} influence of mass fluctuations is reduced. The
coarsening times $\tau_{\text{c}}$ are measured for each trajectory
and finally averaged to determine the mean value for a given set of
parameters $M$ and $p_{\text{left}}$.
Although being simple and fairly robust, this method does provide only
limited insight into the coarsening process.

For the second method we consider the full trajectory of the
relaxation process to estimate the time scale. For each simulation we
record the growth of the largest droplet in terms of its mass
$\overline{M'}(t)$ as a function of simulation time.  The quenched
average of the droplet-growth functions then directly reflects the
state of the relaxation process.  By rescaling $\overline{M'}(t)$ on
the mass as well as the time axis we can then determine the coarsening
time scale.  The mass is rescaled by the expected value of excess in
the steady state $\overline{M'}(t \to \infty)$ and the time is
rescaled by the proposed power law \eqref{eq:scaling}.  If the
proposed scaling law holds, the scaling exponent $\delta$ can be
determined by collapsing the rescaled droplet growth functions for
different sizes of the system to a master curve.  The upper inset in
Fig.~\ref{fig:nucleation-coarsening} shows the result of such a
collapse. To improve stability and error estimation, we perform this
rescaling method over all pairs of trajectories to find the scaling
exponent $\delta$ that yields the best collapse.

\subsection{ZRP}
\label{subsec:results-coarsening-zrp}

We estimate the time scale of the coarsening process for particle
density $\rho=1$ and several values of the hopping asymmetry
$p_{\mathrm{left}}$ to observe the transition of randomly hopping to
driven particles. Here we present our results determined by the method
of first-passage times as well as rescaling the function of the
largest droplet mass $M'(t)$ versus time as those proved to be the
most robust of the considered methods~\footnote{As a cross-check, we
  also rescaled the average droplet count function, to obtain a data
  collapse at the point where only a single droplet would remain.}.
The estimates presented in this section are obtained from $10^{3}$ to
$10^4$ individual simulations of the condensation process for
different system sizes between $100$ and $1000$ sites ($8000$ for
totally asymmetric dynamics) and the various strengths of asymmetric
dynamics $1/2 \le p_{\text{left}}\le 1$.

\begin{figure}
  \centering
  \includegraphics[height=5cm]{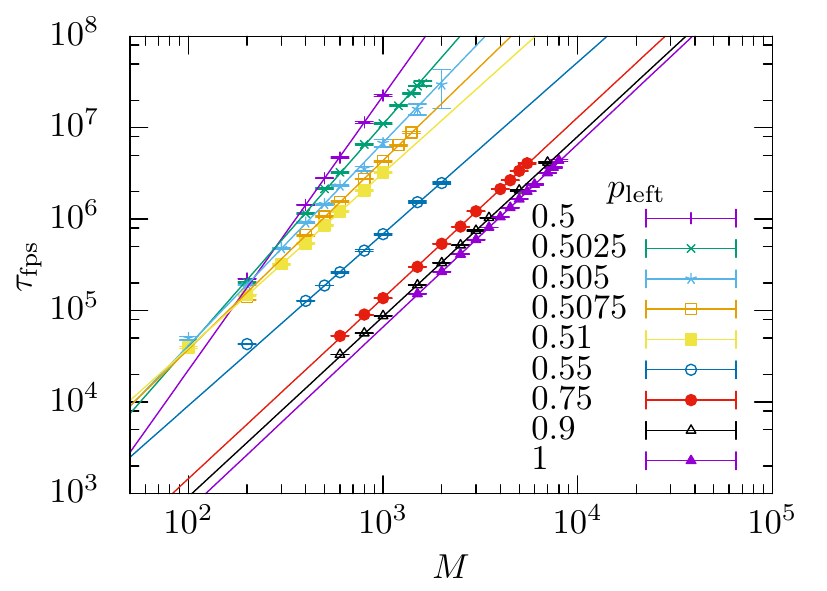}
  \caption{Estimation of the coarsening time-scale exponent
    $\delta_{\text{fps}}$ for the ZRP using the first-passage method
    with scaling ansatz~\eqref{eq:scaling}.}
  \label{fig:zrpfps}
\end{figure}

The coarsening times estimated by the first-passage scaling method are
given in Fig.~\ref{fig:zrpfps}. The scaling exponents of the observed
pronounced power-law dependence of the coarsening time
$\tau_{\text{fps}}$ to the total number of particles $M=\rho L$ are
very close to the value $\delta_{\text{fps}}=1.999\pm0.008$ for
totally asymmetric dynamics for any sufficiently strong partial
asymmetric hopping $p_{\text{left}} > 0.51$. For nearly symmetric
dynamics $0.5\le p_{\text{left}}\le 0.51$, a transition is observed in
the exponent of the coarsening time scale as it grows to
$\delta_{\text{fps,sym}}=2.998\pm0.016$. The specific numerical values
of the determined scaling exponents of the typical coarsening time
$\delta_{\text{fps}}$ are given in Table~\ref{tab:delta-zrp}.

\begin{figure}
  \centering
  \includegraphics[height=5cm]{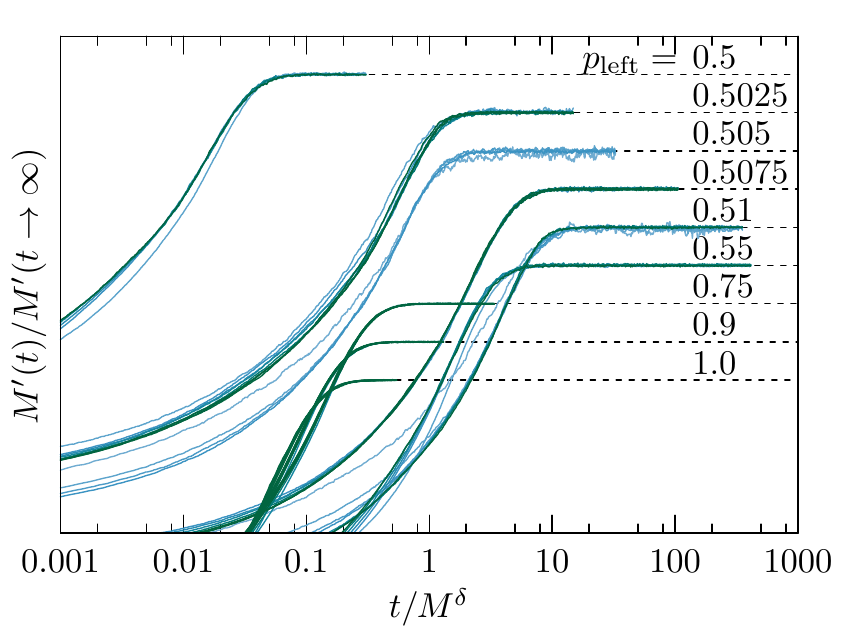}
  \caption{ZRP coarsening: Overview of rescaled largest condensate
    mass trajectories fitted to collapse to master curves in the final
    coarsening dynamics. Groups of curves with identical strength of
    asymmetry are shifted on the vertical axis for readability as
    indicated by the right-hand side labels for the strength of
    asymmetric dynamics $p_{\text{left}}$.  Curves coloured from blue
    to green correspond to time series of smaller (from $L=100$) to
    larger (towards $L=1000$ for symmetric and partially asymmetric
    and $8000$ for totally asymmetric dynamics) systems with overall
    density $\rho=1$. Each curve is computed from $10^4$
    trajectories.}
  \label{fig:zrpmseries}
\end{figure}

\begin{table}
  \caption{Comparison of differently obtained scaling exponents of the
    coarsening time $\delta$ for various strengths of asymmetric dynamics
    $1/2\le p_{\text{left}}\le 1$ with the numerical values of Grosskinsky
    et~al.~\cite{Grosskinsky2003}. To compute these values, we used from
    $10^{3}$ to $10^{4}$ trajectories of systems of size $L=100$ up to
    $1000$ sites (8000 for totally asymmetric hopping) with an overall
    density of $\rho=1$.}
  \begin{tabular*}{\columnwidth}{l@{\extracolsep{\stretch{1}}}*{3}{c}}
    \hline\hline
    $p_{\text{left}}$
    & $\delta_{\text{fps}}$
    & $\delta_{M'}$
    & $\delta_{\text{Grosskinsky2003}}$ \\ 
    \hline
    $1.0$    & $1.999 \pm 0.008$ & $2.0082 \pm 0.0052$ & $1.946 \pm 0.019$ \\
    $0.9$    & $2.089 \pm 0.106$ & $1.993 \pm 0.006$   \\
    $0.75$   & $2.060 \pm 0.098$ & $1.989 \pm 0.005$   \\
    $0.55$   & $1.891 \pm 0.030$ & $2.020 \pm 0.046$   \\
    $0.51$   & $2.035 \pm 0.036$ & $2.035 \pm 0.037$   \\
    $0.5075$ & $2.065 \pm 0.017$ & $2.102 \pm 0.018$   \\
    $0.505$  & $2.197 \pm 0.029$ & $2.369 \pm 0.032$   \\
    $0.5025$ & $2.539 \pm 0.041$ & $2.621 \pm 0.041$   \\ 
    $0.5$    & $2.998 \pm 0.016$ & $3.008 \pm 0.010$   & $2.994 \pm 0.036$ \\
    \hline\hline
  \end{tabular*}
  \label{tab:delta-zrp}
\end{table}

We also determined the coarsening time scale exponent by rescaling the
time series of the largest droplet's mass $M'(t)$.
The obtained rescaled and collapsed data for the ZRP is shown in
Fig.~\ref{fig:zrpmseries}. For any strength of asymmetric hopping, the
curves of the rescaled mass of the largest droplet are collapsed by
minimizing the square differences pair-wise for different system
sizes. To find the coarsening time, this data collapse is performed at
the end of the coarsening stage, when the largest droplet's stationary
mass is approached (see the upper inset for coarsening in
Fig.~\ref{fig:nucleation-coarsening}). The scaling exponents
determined from the good data collapse shown in
Fig.~\ref{fig:zrpmseries} are given in
Table~\ref{tab:delta-zrp}. Additionally we can observe the different
values of the constant prefactors $a$
in the scaling law~\eqref{eq:scaling}. For asymmetric dynamics, the
coarsening time scale is roughly an order of magnitude faster for
strongly asymmetric dynamics $p_{\text{left}}\le
0.75$ compared with weakly asymmetric dynamics. For nearly and fully
symmetric dynamics $p_{\text{left}}>0.5$,
however, the value of this prefactor decreases as the curves shift to
the left and the scaling exponent approaches the value
$\delta_{M'}=3$.
A plot of the coarsening exponents for different strengths of
asymmetric dynamics (see Fig.~\ref{fig:exponents} below) shows a
small transition region for $p_{\text{left}}\gtrsim
0.5$ where the observed scaling exponents change from
$\delta=3$
to $\delta\approx
2$. Finally, comparing our results to the literature (see
Table~\ref{tab:delta-zrp}), we find that our methods work sufficiently
well to proceed to the PFSS model.

\subsection{PFSS}
\label{subsec:results-coarsening-pfss}

\begin{figure}
  \includegraphics[height=5cm]{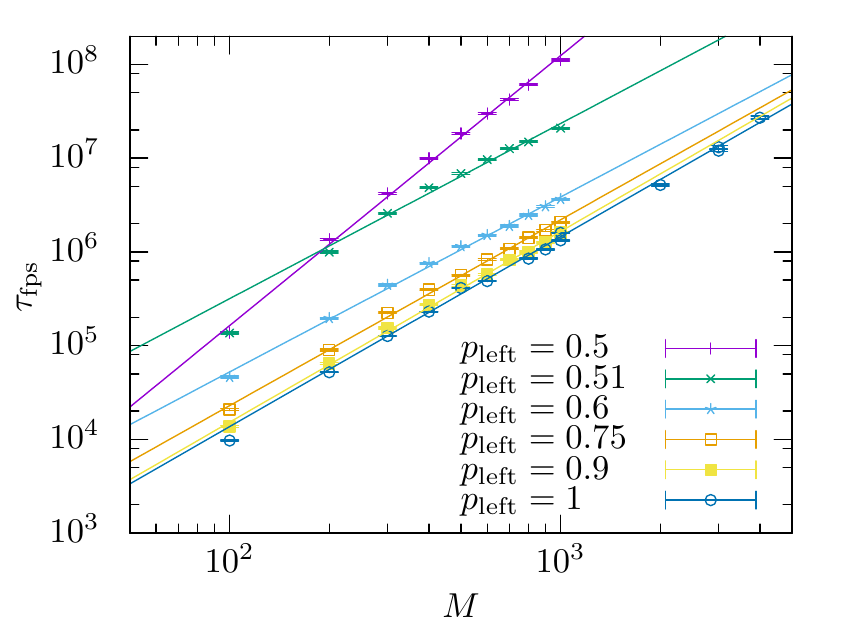}
  \caption{Estimation of the scaling exponent $\delta_{\text{fps}}$ in
    the coarsening regime of the PFSS by fitting the first-passage
    time $\tau_{\text{fps}}$ to the assumed power
    law~\eqref{eq:scaling}. The obtained values are collected in
    Table~\ref{tab:delta-pfss}.}
  \label{fig:pfss-scaling-fps}
\end{figure}

For the PFSS model with short-range interactions we estimate the
exponents of the coarsening time scale $\delta_{\text{fps}}$ using the
first-passage time method and $\delta_{M'}$ by rescaling the time
series of the largest droplet mass as discussed above. Because of the
higher computational cost to simulate this process we limited our
simulations to somewhat smaller system sizes of $L=1000$ sites for
symmetric and partially asymmetric hopping and $L=4000$ for asymmetric
hopping.  Our results for the coarsening times as determined by the
first-passage method are given in Fig.~\ref{fig:pfss-scaling-fps}.
Again, the power-law time scaling is quite pronounced and the scaling
exponents are easily determined as the slope of the curves as
$\delta_{\text{fps}}=2.027\pm 0.040$ for totally asymmetric and
$\delta_{\text{fps}}=2.943\pm 0.070$ for symmetric dynamics.

Here we also determine the numeric values of the scaling prefactor
$a_{\text{fps}}$ to the assumed power law \eqref{eq:scaling} of the
coarsening time scale. The specific values for different strengths of
asymmetric dynamics $p_{\text{left}}$ are given along with the
respective scaling exponents in Table~\ref{tab:delta-pfss}. As for the
ZRP, the constant prefactors increase significantly towards weaker
asymmetric dynamics. We were not able to perform sufficient
simulations for significantly weaker asymmetric dynamics
$0.5<p_{\text{left}}<0.51$ due to the increasing computation costs and
thus did not observe a continuous transition from asymmetric to
symmetric dynamics.

\begin{figure}
  \includegraphics[height=5cm]{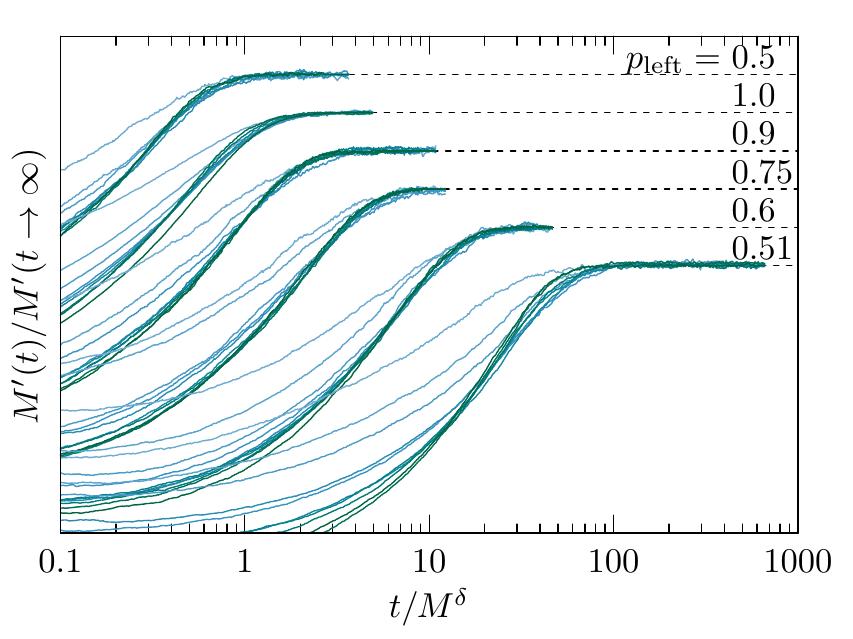}
  \caption{PFSS coarsening: Overview of rescaled largest condensate
    mass trajectories fitted to collapse to master curves in the final
    coarsening dynamics. Groups of curves with identical values of
    $p_{\text{left}}$ are shifted vertically as indicated by the
    right-hand side labels. Curves coloured from blue to green
    correspond to time series of smaller (from $L=100$) to larger
    (towards $L=1000$ for symmetric and partially asymmetric and
    $4000$ for totally asymmetric dynamics) systems. The particle
    density is $\rho=1$ for all systems. Each curve is computed from
    $10^{4}$ trajectories.}
  \label{fig:pfssmseries}
\end{figure}

The rescaled and collapsed curves determined using our second method
are given in Fig.~\ref{fig:pfssmseries} and yield coarsening exponents
$\delta_{M'}=2.005\pm 0.038$ and $\delta_{M'}=2.855\pm0.031$ for
totally asymmetric and symmetric dynamics, respectively. The full set
of exponents for different strengths of partially asymmetric dynamics
$p_{\text{left}}$ is given in Table~\ref{tab:delta-pfss}. For the PFSS
model the collapse shown in Fig.~\ref{fig:pfssmseries} is less
distinct for the smaller system sizes of $L=100$ sites than for the
larger sizes $L\ge 200$.  Although this rescaling process to collapse
curves and determine the exponent by the rescaling parameter is rather
involved this method yields much more stable and somewhat more robust
results than the first-passage method. This is a result of its
implementation using pair-wise collapse of the individual trajectories
of different system sizes and global minimization of the total error
function of the data collapse in the final coarsening stage.

\begin{figure}
  \includegraphics[height=5cm]{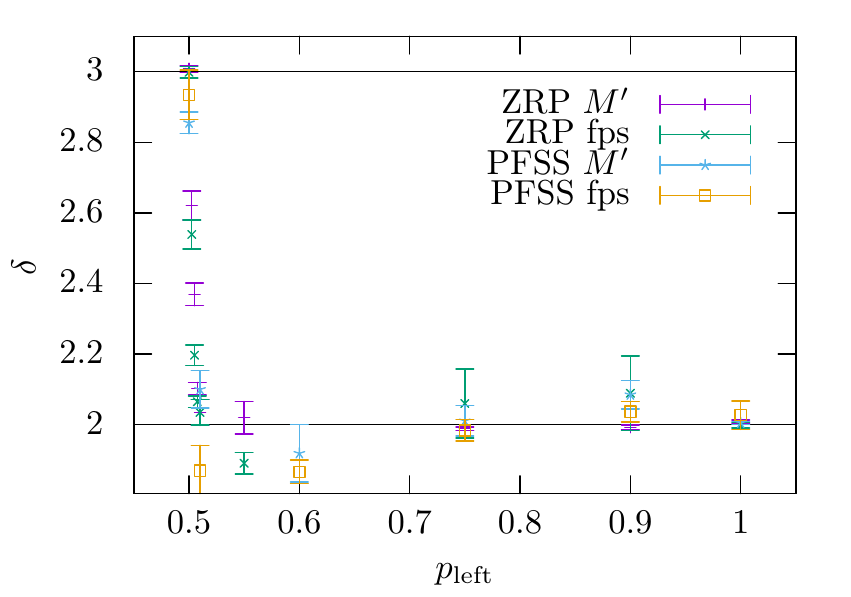}
  \caption{Scaling exponents  $\delta$ of the coarsening  times versus
    strength of asymmetric dynamics $p_{\text{left}}$ for both ZRP and
    PFSS transport processes.}
  \label{fig:exponents}
\end{figure}

\begin{table}
  \caption{%
    Exponents $\delta_{\text{fps}}$ and $\delta_{M'}$ of the condensation 
    time scale with prefactor $a_{\text{fps}}$ for the PFSS transport 
    model as in \eqref{eq:scaling} determined by first-passage scaling 
    ($\delta_{\text{fps}}$) as well as pair-wise rescaling ($\delta_{M'}$)
    of the largest condensate mass, respectively. Values are computed
    from $10^4$ trajectories of systems of size
    $L=100$ up to $1000$ sites ($4000$ for totally asymmetric hopping)
    with an overall density of $\rho=1$. Simulations of the condensation
    process for higher densities  $\rho=2, 3$ yield quantitatively the 
    same results.}
  \sisetup{}
  \begin{tabular*}{\columnwidth}{l@{\extracolsep{\stretch{1}}}*{3}c}
    \hline\hline
    $p_{\text{left}}$ &
    $\delta_{\text{fps}}$ &
    $\delta_{M'}$ &
    $a_{\mathrm{fps}}$
    \\
    \hline
    \(1.0\) & $2.027 \pm 0.040$  & $2.005 \pm 0.038$ & $1.21 \pm 0.31$ \\
    $0.9$   & $2.037 \pm 0.029$  & $2.085 \pm 0.040$ & $1.28 \pm 0.24$ \\
    $0.75$  & $1.984 \pm 0.030$  & $2.011 \pm 0.043$ & $2.45 \pm 0.47$ \\
    $0.6$   & $1.867 \pm 0.033$  & $1.919 \pm 0.081$ & $9.63 \pm 2.00$ \\
    $0.51$  & $1.870 \pm 0.072$  & $2.100 \pm 0.053$ & $58 \pm 26$ \\ 
    $0.5$   & $2.934 \pm 0.070$  & $2.855 \pm 0.031$ & $\hphantom{0}0.28 \pm 0.090$ \\
    \hline\hline
  \end{tabular*}
  \label{tab:delta-pfss}
\end{table}

Table~\ref{tab:delta-pfss} and Fig.~\ref{fig:exponents} summarize the
results of the scaling exponents as well as the scaling prefactor
$a_{\text{fps}}$ determined by the first method.

\section{Conclusions}
\label{sec:summary}

In this work we considered the dynamics of the two stages, nucleation
and coarsening, of the condensation process in the zero-range process
(ZRP) condensation model as well as a stochastic transport process
with pair-factorized steady states (PFSS) and condensation
dynamics. To obtain the typical time scales of the nucleation and
coarsening process, we considered the power-law scaling observed for
zero-range processes with condensation dynamics and employed
complementary numerical methods.

For the nucleation regime in both processes we find most notably the
strong dependence of the scaling exponent of the nucleation time on
the choice of the typical droplet mass that acts as a marker to the
end of the regime. The determined exponents for the ZRP for the three
choices of the typical droplet mass, for totally asymmetric and
symmetric dynamics respectively, are $\delta_{\text{ZRP,nucl}}=2$ and $3$ for
$m_{\text{t,lin}}\propto(\rho-\rho_{\text{c}})L$ linear in the excess
mass, $\delta_{\text{ZRP,nucl}}=1$ and $3/2$ for
$m_{\text{t,sqrt}}\propto\sqrt{(\rho-\rho_{\text{c}})L}$ linear in the
square root of the excess mass and $\delta_{\text{ZRP,nucl}}=0$ for
$m_{\text{t,const}}$ independent of the system size. These exponents
were also expected using a simple heuristic approximation of the
involved time scales.  
The almost perfect data collapse of the droplet-count function for the
threshold based on our approximation of the critical droplet size for
the ZRP suggests that indeed the assumption of a typical droplet mass
$m_{\text{t,const}}$ that is an intensive variable, i.e.\ does not
depend on the system size, is physical for the ZRP and the PFSS model
as well. However, considering the versatility of these models allowing
a large variety of mappings to other models, the freedom to choose the
type of the droplet mass threshold with respect to such a mapping and
obtain an appropriate nucleation time scale seems to be of advantage.

For the pair-factorized steady states model we determined scaling
exponents using the same types of typical droplet masses with
differences in the prefactors to account for the larger extended
droplets. Remarkably, our obtained values of these exponents directly
correspond to those of the ZRP plus a shift of $1/2$ towards longer
nucleation times for the considered droplet mass thresholds that
depend on the system size. The specific values, for totally asymmetric
and symmetric dynamics, are $\delta_{\text{PFSS,nucl}}=5/2$ and $7/2$
for the threshold $m_{\text{t,lin}}$ linear in the excess mass,
$\delta_{\text{PFSS,nucl}}=3/2$ and $2$ for $m_{\text{t,sqrt}}$ linear
in the square root of the excess mass and
$\delta_{\text{PFSS,nucl}}=0$ for the threshold $m_{\text{t,const}}$
independent of the system size.

It would be rewarding for further research to study the details of the
involved nucleation processes that lead to the observed relation
between the exponents for the PFSS and those of the ZRP. Furthermore,
it is worth studying the crossover to the coarsening regime in
stochastic transport processes in more detail.

For the coarsening time scale exponents we determined results
employing two independent numerical methods. To test these, we first
reproduced the scaling exponents of the ZRP determined analytically
and numerically by Grosskinsky et al.~\cite{Grosskinsky2003} for
symmetric as well as totally asymmetric dynamics. Our results for the
scaling exponents of the coarsening times are
$\delta_{\text{ZRP,tas}}=2.0082 \pm 0.0052$ and
$\delta_{\text{ZRP,sym}}=3.008 \pm 0.010$  as well as
$\delta_{\text{PFSS,tas}}=2.005 \pm 0.038$ and
$\delta_{\text{PFSS,sym}}=2.855 \pm 0.031$ for the PFSS model, both
for symmetric and totally asymmetric dynamics respectively. In
contrast to the nucleation regime, the coarsening exponents of the ZRP
and PFSS models match very well suggesting that, although the early
condensation dynamics differ, the coarsening process in these
processes is much more similar.
We supplemented these values with results for partially asymmetric
dynamics and showed the transition in the observed coarsening times
from symmetric to predominantly asymmetric dynamics at
$p_{\text{left}} \ge 0.51$ from $\delta=3$ to $2$ accompanied by a
distinct change in the prefactor of the scaling. This is most likely
due to the finite system size. For the PFSS in the coarsening regime
we discussed results for the scaling exponents with symmetric, as well
as partially and totally asymmetric dynamics. We were not able to
directly observe such a transition in the coarsening exponents from
$\delta=3$ to $2$ for very weakly asymmetric dynamics as for the ZRP,
due to the increase in the computational costs for such
parametrisations. The prefactor of the scaling, however, rapidly
increases for nearly symmetric dynamics as observed for the ZRP as
well.

\acknowledgements{We are grateful to Johannes Zierenberg for
  insightful suggestions and discussions of the nucleation dynamics.
  We thank the DFG (German Science Foundation) for financial support
  under Grant No.\ JA 483/27-1.  We further acknowledge support by the
  DFH-UFA Doctoral College ``$\mathbb{L}^4$'' under Grant No.\
  CDFA-02-07 as well as the EU through the Marie Curie IRSES network
  DIONICOS under contract No.~PIRSES-GA-2013-612707.}


%

\end{document}